\documentclass[twocolumn,amsmath,amssymb,prl,superscriptaddress]{revtex4-1}

\usepackage{graphicx}
\usepackage{dcolumn}
\usepackage{bm,epsfig}

\begin{document}

\newcommand{\Ce}{Ce$_2$RhIn$_8$}
\newcommand{\La}{La$_2$RhIn$_8$}
\newcommand{\CRI}{CeRhIn$_5$}
\newcommand{\Ef}{$E_F$}
\newcommand{\kf}{$\textbf{k}_F$}

\title{Fermi surface collapse and energy scales in Ce$_{2}$RhIn$_{8}$}

\author{F. Rodolakis} 
\affiliation{Advanced Photon Source, Argonne National Laboratory, Argonne, IL 60439, USA}
\author{C. Adriano}
\affiliation{Department of Physics, University of Illinois at Chicago, Chicago,IL 60607, USA}
\affiliation{Instituto de F\'isica \lq\lq Gleb Wataghin\rq\rq,UNICAMP, Campinas-SP, 13083-970, Brazil}
\author{F. Restrepo}
\affiliation{Department of Physics, University of Illinois at Chicago, Chicago,IL 60607, USA}
\author{P. F. S. Rosa}
\affiliation{Instituto de F\'isica \lq\lq Gleb Wataghin\rq\rq,UNICAMP, Campinas-SP, 13083-970, Brazil}
\affiliation{Condensed Matter and Magnet Science, Los Alamos National Laboratory, Los Alamos, NM 87545, USA}
\author{P. G. Pagliuso}
\affiliation{Instituto de F\'isica \lq\lq Gleb Wataghin\rq\rq,UNICAMP, Campinas-SP, 13083-970, Brazil}
\author{J. C. Campuzano}
\affiliation{Department of Physics, University of Illinois at Chicago, Chicago,IL 60607, USA}

\date{\today}

\begin{abstract}
In some metals containing a sub-lattice of rare earth or actinide ions, free local  $f$ spins at high temperatures  dissolve into the sea of quantum conduction electrons at low temperatures, where they become mobile excitations. Once mobile, the spins acquire charge, forming electrons of heavy mass, known as heavy fermions.
In turn, the incorporation of heavy charges into the conduction sea  leads to an increase in the volume of the Fermi surface.
This process, called Kondo scattering, is accompanied by a dramatic, temperature dependent transformation of the electronic interactions and masses. Since the Kondo phenomena is controlled by quantum fluctuations, here we ask, at which point does the Fermi surface change character?  A priori, the answer is not clear, since near its onset, the Kondo effect cannot be described as a simple hybridization of electronic eigenstates. Conventional descriptions of this Kondo scattering process consider that hybridization, Fermi volume change, and $f$-electron mobility occur simultaneously.  However, using angle resolved photoemission spectroscopy to measure the evolution of excitations, we find that the changes of the Fermi surface emerge at temperatures an order of magnitude higher than the opening of the hybridization gap, and two orders of magnitude higher than the onset of the coherent character of the $f$-electrons. We suggest that the large changes in Fermi volume, driven by electronic fluctuations, occur at temperatures where the various $\Gamma_x \to \Gamma_y$ crystal field-split $f$ levels  become accessible to conduction states of the corresponding symmetries. The separation of these energy scales significantly modifies the conventional description of the Kondo lattice effect, which still lacks a full theoretical description. 
\end{abstract}


\maketitle

\begin{figure}
\centerline{\includegraphics[width=.5\textwidth]{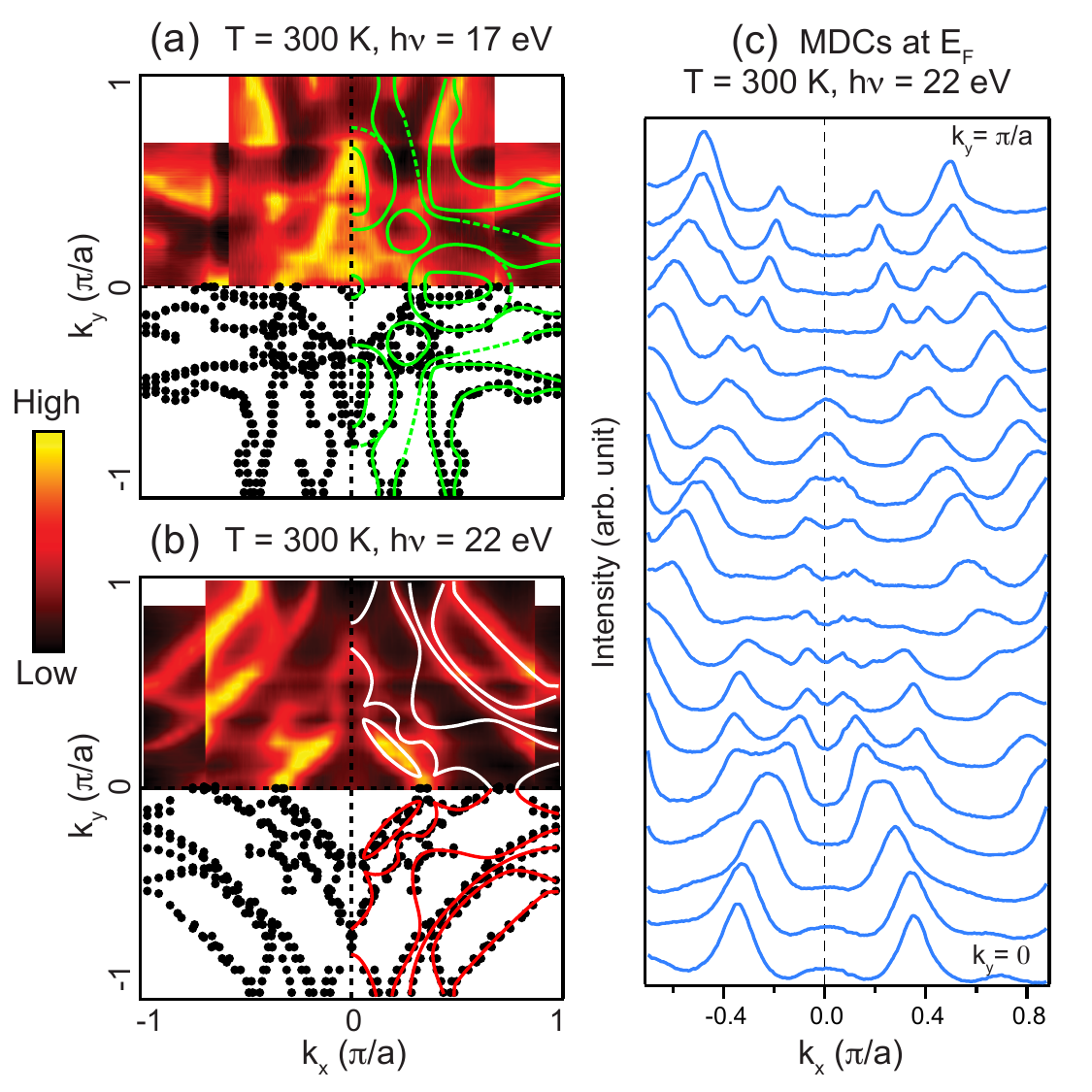}}
\caption{Fermi Surfaces at room temperature in \Ce. (a,b) Symmetrized ARPES intensity maps as a function of $k_x$ and $k_y$ (top halves) measured at 300K and integrated over an energy windows of 20~meV centered at 0 binding energy (\Ef): (a) $h\nu=$17~eV, (b) $h\nu=$22~eV. The bottom halves show the \kf\ values extracted from the corresponding MDCs at \Ef\ (see Supplemental Material). Continuous lines following those dots are superimposed on right halves (dashed lines in (a) are guesses based on comparison with theory).
(c) MDCs at  0 binding energy ($T=300$~K, $h\nu=22$~eV),  for different angles ranging from $\theta = 0^\circ$ (i.e. $k_y=0$) to $\theta = 18^\circ$ (i.e. $k_y=\pi/a$).}
\label{fig:Fig300K}
\end{figure}

\begin{figure}
\centerline{\includegraphics[width=.5\textwidth]{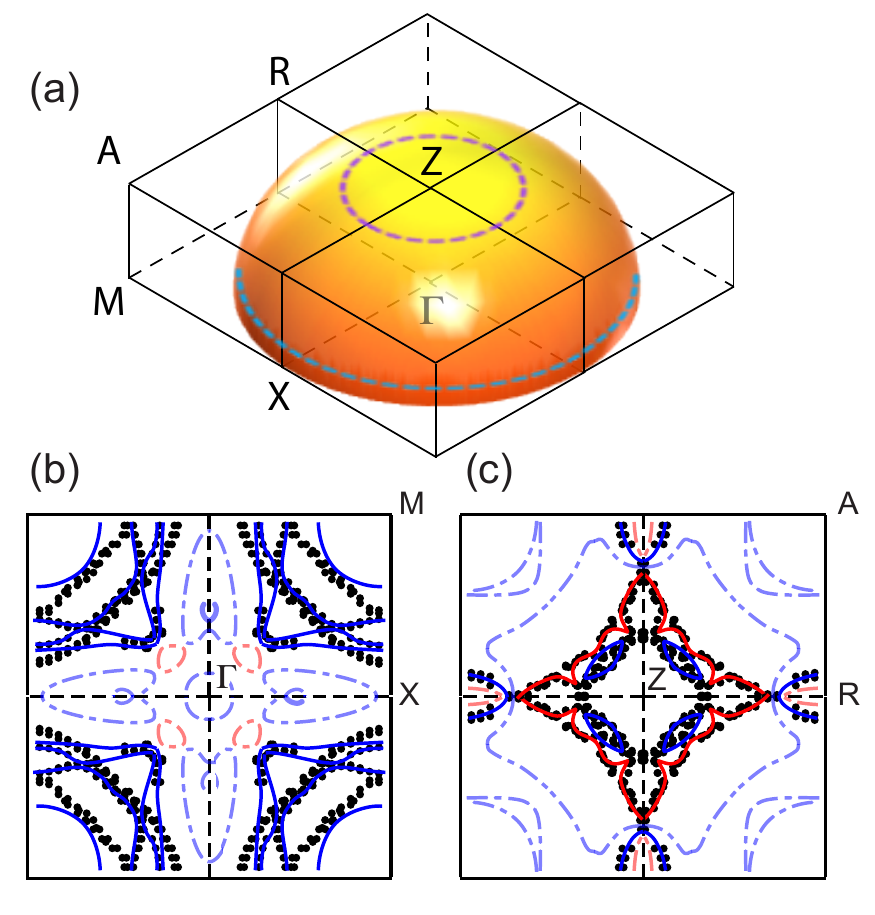}}
\caption{Comparison with calculated Fermi Surfaces of \La. 
In three-dimensional samples, ARPES data at different $k_{//}$ (\emph{i.e.} the momentum parallel to the surface) correspond to different values of $k_z$ which fall on a spherical surface, as sketched in the top panel (a). Therefore, points at a given radial distance from the origin in ARPES intensity maps (Fig.~\ref{fig:Fig300K}) correspond to cuts of the spherical surface as indicated by te dashed circles  in (a).
The calculated Fermi surfaces of \La\ in the $\Gamma XMX$ and $ZRAR$ planes from Refs.~\cite{Ueda2004,RAJ2005} are shown in panel (b) and (c) respectively (lines). The experimental \textbf{k}$_F$ values  measured in \Ce\ at room temperature and $h\nu=$22~eV are plotted for comparison (dots). Solid lines emphasize the correspondence between theory and experiment. Electron and hole pockets are displayed in blue and red, respectively.
}
\label{fig:FigTheory}
\end{figure}

Ce$_{2}$RhIn$_{8}$ has a three-dimensional Fermi surface~\cite{RAJ2005}: in Fig.~\ref{fig:Fig300K} we show ARPES intensity maps measured at 300~K for photon energies $h\nu$ of (a) 17~eV, and (b) 22~eV, respectively. In three-dimensional samples, points at different radial distances from the origin in plots (a) and (b) correspond to different values of $k_z$, the momentum perpendicular to the surface, as illustrated in Fig.~\ref{fig:FigTheory}(a). At the Fermi energy, the relationship is given by $k_z=\frac{1}{\hbar} \sqrt{2m (E_{kin} cos^2{\theta}+V_0)}$, where $E_{kin}=h\nu-\phi$, with $\phi$ the work function, $\theta$ the photoelectron emission angle, and $V_0$ the inner potential which refracts the photoelectron upon leaving the sample. Ideally, one would like to maintain constant $k_z$ for all $k_x, k_y$. However time requirements make such experiments impractical. Instead, using the dispersion of states at $k_x=\pm0.21(\pi/a)$ and $k_y=0$ as a function of photon energy $h\nu$, we determine that $V_0 =12 \pm 1$~eV (see Supplemental Material). This allows us to place each measurement point at the correct \textbf{k}. More importantly, meaningful conclusions about changes in the Fermi volume are still possible, since we are measuring at fixed $|\bf{k}|$ as a function of temperature.

Although the ARPES intensity maps in Fig.~\ref{fig:Fig300K} provide an appealing visual overview, they do not yield unambiguous spectroscopic information. Instead, we use the momentum distribution curves (MDCs), which are plots of the photoemission intensity at a fixed binding energy ($E_B$) vs. momentum \textbf{k}, as shown in Fig.~\ref{fig:Fig300K}(c). The peaks in the MDCs at the Fermi energy \Ef, indicated by the dots in the bottom halves of Figs.~\ref{fig:Fig300K}(a,b), yield the Fermi surface (FS). Continuous curves that follow the points are drawn on top of the intensity maps on the right halves of Figs.~\ref{fig:Fig300K}(a,b). 
The origin of the intensity plots shown in Fig.~\ref{fig:Fig300K} correspond approximatively to $\Gamma_5$ (the center of the 5th BZ) for $h\nu=17$~eV and Z$_5$ for $h\nu=22$~eV (b).

Surface effects are known to be problematic for the interpretation of ARPES data of heavy fermion materials~\cite{ALLEN2005}. We employ the following strategy to assure ourselves that none of the features shown in Fig.~\ref{fig:Fig300K} originate from surface states. It is reasonable to assume that the $f$-electrons are localized at room temperature~\cite{HEWSON1993,STEWART1984,LOHNEYSEN2007}. We may therefore compare the measured Fermi surface to LDA calculations performed in \La\ by Ueda \emph{et al.} \cite{Ueda2004,RAJ2005}. \La\  has the same structure as \Ce\  studied here, except that the replacement of Ce by La removes the $f$-states from the calculation.  If the measured FS sheets show a reasonable approximation to the calculations, such as the same number of sheets with a similar dispersion, we can safely assume that the states in question are not surface states. In Fig.~\ref{fig:FigTheory} we show the experimental points measured at $T = 300$~K and $h\nu=22$~eV (black dots) from Fig.~\ref{fig:Fig300K}. (b) shows the experimental data around the Brillouin zone (BZ) corners compared to the calculations for the ($\Gamma$XMX) plane (lines), and in (c), experimental dots in the vicinity of the BZ center are compared to calculations on the (ZRAR) plane (lines). Electron and hole pockets are displayed in blue and red, respectively.

We note that it is more difficult to compare calculation and experiments at 17~eV photon energy due to the extremely complicated topology of the hole-like FS at the center of the BZ \cite{Ueda2004}. Instead, we compare the experimental results to calculations using 22~eV photon energy data.

\begin{figure*}[ht]
\begin{center}
\centerline{\includegraphics[width=.9\textwidth]{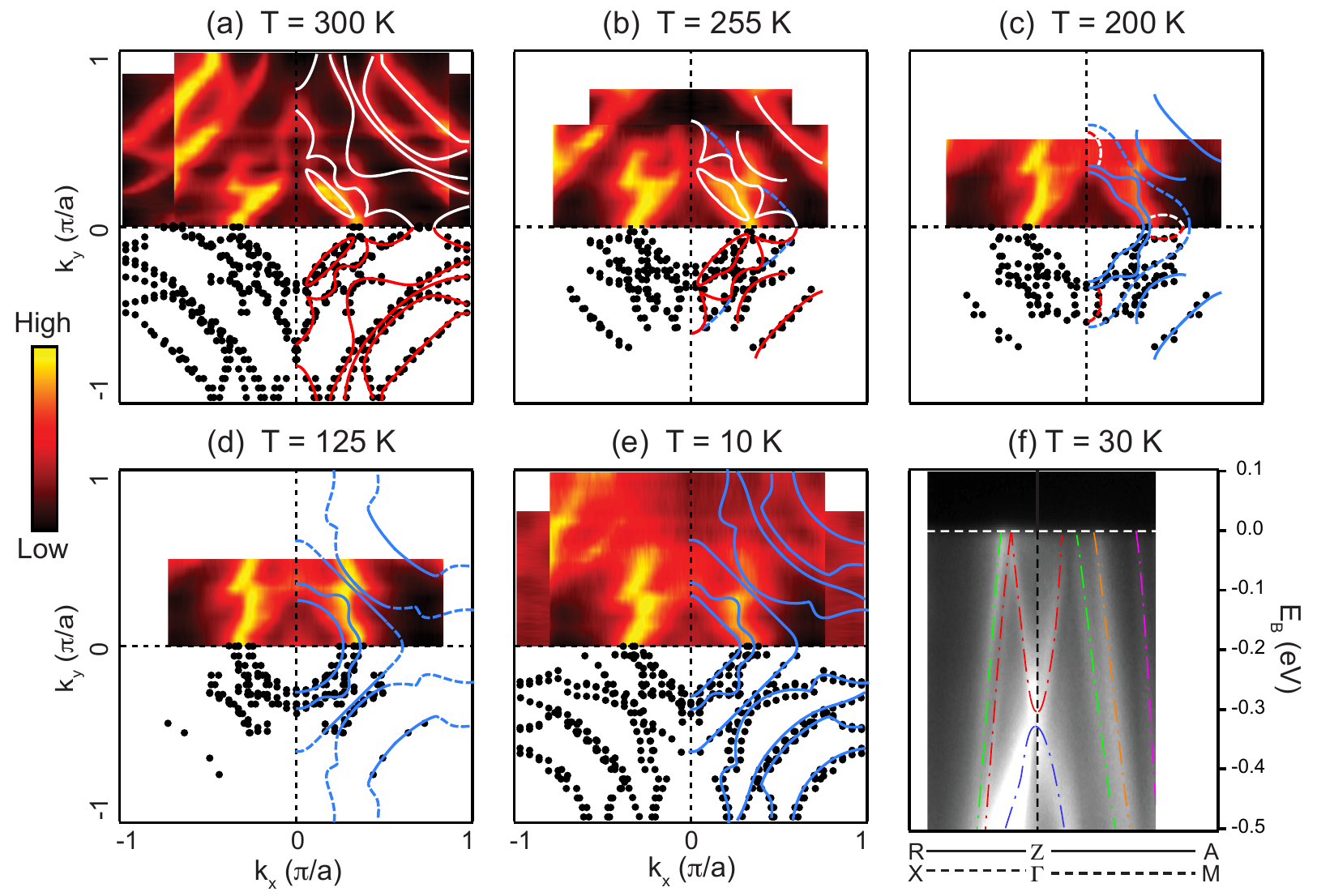}}
\caption{Fermi Surfaces evolution with temperature in \Ce. Symmetrized ARPES intensity maps integrated over an energy windows of 20~meV centered at 0 binding energy (\Ef) as a function of $k_x$ and $k_y$ ($h\nu=22$~eV): (a) T = 300~K, (b) T=255~K, (c) 200~K, (d) T=125~K and (e) T=10~K. Colored lines shows the FSs determined from Fermi vector analysis. (f) Experimental band dispersion in \Ce: ARPES intensity maps as a function of wave vector and binding energy measured at $h\nu=22$~eV along high symmetry directions at T=30~K.  Grey and white areas correspond to the band and colored lines are guides to the eye. Note that some bands are not visible here due to strong matrix elements effects (see Supplemental Material).}\label{fig:FigFST}
\end{center}
\end{figure*}

The correspondence of the calculations to different regions of our plots, depending on the high symmetry plane and direction considered, are made explicit in the  sketch shown in Fig.~\ref{fig:FigTheory}(a) (see Supplemental Material for a more quantitative analysis). The correspondence of each measured Fermi surface sheet to one in the calculation allows us to conclude that none of the experimental data shown and discussed here originates from surface states. In summary, at room temperature for $h\nu= 22$~eV, we observe 4 cylindrical but corrugated electron-like FS sheets centered at the BZ corner, and a large "star-like" hole sheet enclosing 4 small electron-pockets. 

Having understood the high temperature state with asymptotically free $f$ moments, we can now examine the evolution of the electronic excitations with decreasing temperature, shown in Fig.~\ref{fig:FigFST}. Comparing data measured at 300K (Fig.~\ref{fig:FigFST}(a)) with that measured at 10K (Fig.~\ref{fig:FigFST}(e)), one can see that the three outermost barrels mainly undergo a slight distortion. In contrast, states around the BZ center exhibit dramatic changes in their topology at low temperatures. A clear dichotomy in behavior has set in: the states mostly parallel to the [100] direction, which we call in-plane (\textit{i.e.} mostly 2$p$ non-dispersing) states show little mixing with the $f$ levels, while the out-of-plane ones show the largest mixing~\cite{ADRIANO2013}. We arrive at this conclusion because the out-of-plane states around the center of the BZ show the largest dispersion along $k_z$ \cite{Kaminski2014}.

The main changes going from high to low temperatures are the disappearance of the small pockets, and the splitting of the large star-like hole-FS observed at high temperature into two electron-like sheets, which can also be seen from the band dispersion at 30~K along the high symmetry lines shown in Fig.~\ref{fig:FigFST} (f).  These changes lead to a collapse of the Fermi surface volume at high temperatures as it looses one electron. Note that between $T = 30$~K and 10~K the peak of the spectral function shifts, as one would expect from the presence of a gap, but a substantial amount of spectral weight remains at the chemical potential and there is no noticeable change in the band dispersion~\cite{ADRIANO2013}. Due to the strong fluctuations and correlations (i.e. large spectral widths), and small energy shifts of the hybridized states, the energy gap at 10~K is not apparent in the MDC's~\cite{Norman2001}. As a  consequence, the low temperature BZ remains fully occupied, and the mixing can now be described as a standard  hybridization. 

Only once the temperature increases above 125~K does the volume collapse. Indeed, when increasing the temperature from 10~K to 125K, (Fig.~\ref{fig:FigFST}(d)), well above the spectral gap closure, only very small changes can be observed compared to the more dramatic changes occurring above 200~K - \emph{i.e.} about 170~K above the hybridization gap closure~\cite{ADRIANO2013}. The change occurs mostly as the band located along the [010] direction $\big((k_x,k_y)\approx(0,1/2)(\pi/a)\big)$ starts splitting to form the petals of the "star-like" high temperature FS sheet.

A detailed MDC analysis (described in the supplemental material) indicates that at intermediate temperatures, for example 255~K (Fig.~\ref{fig:FigFST}(b)), the FS can be described as coexisting sheets from low temperature states together with sheets originating from room temperature states -- although we note that more than one domain may be present. Nonetheless, we must still conclude that FS sheets of different symmetries mix with the crystal field-split $f$ level of appropriate symmetry and energy at different temperatures.

From extensive temperature dependence measurements, we estimate that the crossover is centered at about $\sim$200~K, which coincides with the minimum of the resistivity curve of \Ce\  with temperature~\cite{ADRIANO2010}. Near this temperature, the fluctuations have become sufficiently large for the Kondo scattering to overcome the phonon scattering, causing the resistivity to increase with further decrease in temperature. In CeIrIn$_5$, Choi \emph{et al.}~\cite{Choi2012} predict that the onset of the crossover where the small FS begins distorting towards the low temperature FS occurs at 130~K, while the composite quasiparticles only emerge below 50~K. Experimental data in  CeRh$_{1-x}$Co$_x$In$_5$ also shows the occurrence of FS reconstruction much earlier than the quantum critical transition~\cite{Goh2008}.

Our data clearly show that the change in the Fermi volume must be thought of as fluctuation driven, since the incorporation of the $f$-electrons into the Fermi volume occurs at temperatures one order of magnitude higher than the opening of the conventional hybridization gap. Even though one might be tempted to call it hybridization, since strictly speaking hybridization is a \emph{local} phenomena, the states are strongly fluctuating in time, quite unlike conventional hybridization. Once the temperature is sufficiently low (by nearly an order of magnitude), and the fluctuations of the conduction electrons have been sufficiently quenched, a spectral gap appears, as shown in Ref. \cite{ADRIANO2013}. There are three clearly distinct crossover temperatures: one where Kondo scattering begins to mix the different $f^1$ levels with the conduction electron sea, and another where the fluctuations have been sufficiently quenched for a spectral gap to appear, and finally the onset of the coherence at very low temperature ($T\sim5$~K).  

Indeed, the crystalline-electric-field (CEF) level scheme for \Ce\ is composed of two magnetic doublets split from a doublet ground state by 70~K and 195~K, respectively~\cite{Malinowski2003}. The lowest two doublets have the same symmetry,  while the one at 195~K has not. 
Therefore, the observation of the gradual collapse of the Fermi volume brings information about the CEF into the character of the Fermi surface. It is not surprising then that small changes in composition~\cite{ADRIANO2013}, and/or applied pressure leads to significant changes in the electronic states.
In fact, a subtle evolution of the CEF effects have been claimed to be an important parameter to tune the ground state from antiferromagnetic to superconducting in this class of materials~\cite{PAGLIUSO2002b,CHRISTIANSON2004,PAGLIUSO2006,KUBO2006,KUBO2007,WILLERS2010,FLINT2010,FLINT2008,GEORGES2014,WILLERS2013}. The results reported here showing a direct correlation between the CEF excitations and FS effects, even at high temperature, can bring important new insights to the understanding of how this tuning may occur.


\begin{acknowledgments}
This work was supported by the Center for Emergent Superconductivity, an Energy Frontier Research Center funded by the US Department of Energy, Office of Science, Office of Basic Energy Sciences under Award Number DE-AC0298CH1088. Work performed at the Synchrotron Radiation Center, University of Wisconsin (Award No. DMR-0537588) (F.R.S., F.R., and J.C.C).  C.A., P.F.S.R. and P.G.P. thank FAPESP (in particular grants No 2006/60440-0, 2009/09247-3, 2010/11949-3, 2011/01564-0, 2011/23650-5, 2012/04870- 7), CNPq, and FINEP-Brazil for supporting this work.
\end{acknowledgments}


\section{Supplemental Material.}

\subsection{Experimental details.}

The ARPES data were obtained on the U1 4m-NIM beamline at the Synchrotron Radiation Center in
Wisconsin using a Scienta R4000 spectrometer. The measurements
were performed at a pressure of 2 to 5 x 10$^{-11}$ Torr and a clean
surface was obtained by \textit{in situ} cleavage of
the crystal perpendicular to the [001] direction. The selected
energies were 17 and 22~eV, with the polarization parallel to the [100] direction. The energy resolution ranged from 13 meV at low temperature to 25 meV at room temperature.
The photon energy dependence detailed below was carried out in similar conditions on the U9 VLS-PGM beamline using a Scienta 200U analyzer with an energy revolution of about 25~meV. 
The single crystals were grown
using In-flux method~\cite{Pagliuso2001}; crystals structures and
phase purity were checked by x-ray powder diffraction. 

\subsection{Data analysis.}

\begin{figure}
\makeatletter 
\renewcommand{\thefigure}{S\@arabic\c@figure}
\makeatother
\centerline{\includegraphics[width=.5\textwidth]{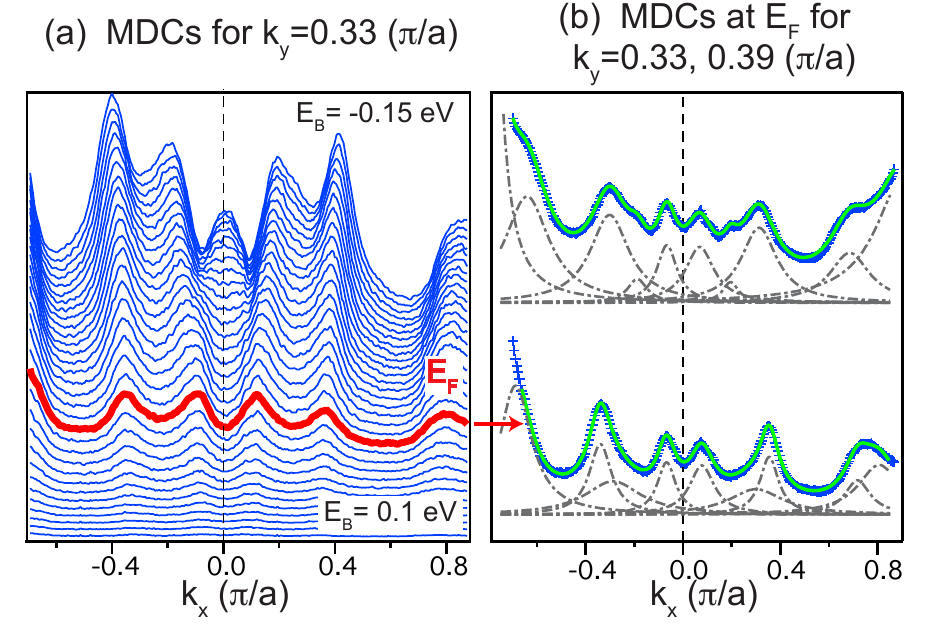}}
\caption{Extraction of the Fermi surfaces sheets from MDCs: (a) shows the MDCs at room temperature for different binding energies near $E_F$ at $k_y = 0.33  (\pi/a)$. The thick red line corresponds to the MDC at $E_F$. (b) Fit of the MDCs at \Ef\  for 2 consecutive angles ($\theta = 6^{\circ}$ and $\theta = 7^{\circ}$) at room temperature and $h\nu=22$~eV. Blue crosses are experimental data, grey dashed lines are the different lorentzians composing the corresponding fit function shown by the green solid lines. }
\label{fig:FigMDC}
\end{figure}

To determine accurately the different FS sheets in \Ce, we used the momentum distribution curves (MDC), \emph{i.e.} plots of the photoemission intensity as a function of momentum for a fixed binding energy, an example of which is shown in Fig.~\ref{fig:FigMDC} (a). By using a sum of lorentzians~\cite{Norman2001} to fit the MDCs at \Ef\ for all the angles covering the BZ (Fig.1(c)), we can follow the band dispersion and obtain a plot where each dot corresponds to a different Fermi vector \kf\ (see bottom halves of Fig.~1(a,b) and Fig.~3(a-e)). Examples of the fitting procedure are displayed for 2 consecutive angles in Fig.~\ref{fig:FigMDC}(b).

Note that the results are not exactly symmetric. First, by comparing left and right sides ($\pm k_x$), the peak positions are not exactly identical: the dots are not perfectly overlapping. Besides the resolution, that can be due to a very slight sample misalignment. Also, as matrix elements are very strong in \Ce, even a very  small misalignment can introduce a noticeable effect on the measurement: some bands which are not visible due to destructive matrix elements can appear when the measurement is perform slightly away from the mirror plane. To take those effects into account, both sides are 4-fold symmetrized, which give us an error bar on the actual \kf\ positions. 

There is a strong matrix element dependance on the direction of the measurement ($\pm k_y$), \emph{i.e. }changing the measuring angle $\theta$ by going away or towards the beam. Some bands are only visible on one specific side. Both measurement were used to get a complete picture of the FS. The 4-fold symmetry of the data has been checked by measuring the same sample before and after a $90^\circ$ rotation, which give the exact same result.

From the \kf\ positions, combined with a careful band dispersion analysis, we can then identify the different FS sheets which are indicated by the colored curves in Fig.~1(a,b) and Fig.~3(a-e).


\subsection{Photon energy dependence.}
\begin{figure}
\makeatletter 
\renewcommand{\thefigure}{S\@arabic\c@figure}
\makeatother
\centerline{\includegraphics[width=.5\textwidth]{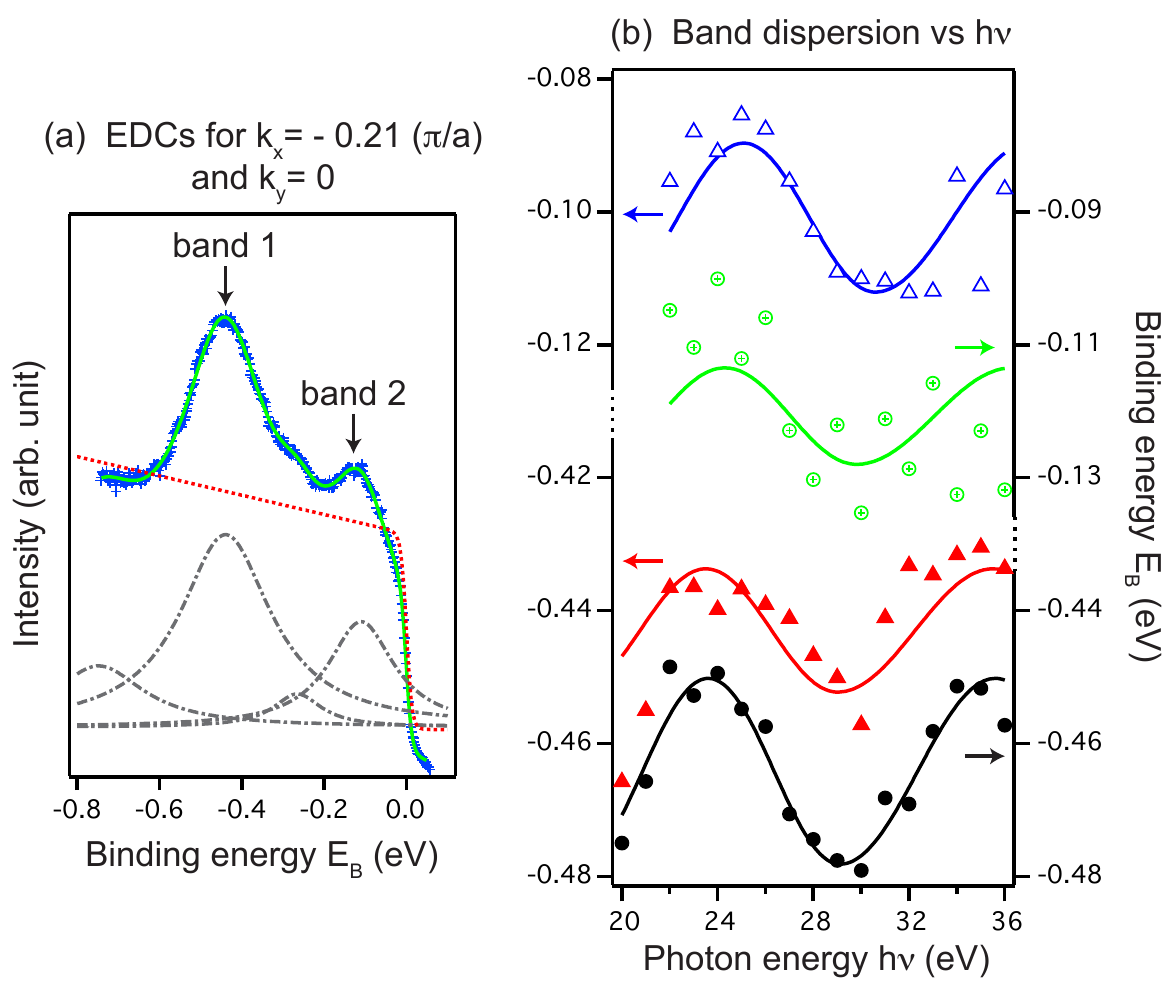}}
\caption{Photon energy dependence: (a) Fit of the EDC for $h\nu=23$~eV, $k_x=-0.21(\pi/a)$ and $k_y=0$. Blue crosses are experimental data, solid green line is the corresponding fit composed by a sum of lorentzians (grey dashed lines) multiplied by the Fermi function (red dotted line). (b) Band dispersions as a function of the photon energy $h\nu$ for bands 1 (solid red triangles) and 2 (open blue triangles) at $k_x=-0.21(\pi/a)$ using the fitting procedure described on panel (a). The green crossed circles show the result obtained with the symmetric of band 2 ($k_x=+0.21(\pi/a)$). The black dots show the band dispersion using a simple gaussian fit for band 1. Solid lines show the fits obtained using the function $V(x)$.
}
\label{fig:FigV0}
\end{figure}

We have studied the photon energy dependence in order to estimate the value of the  inner potential $V_0$ in \Ce. Measurement were carried out at low temperature ($T=20$~K) along the [100] direction at $k_y=0$ using photon energies ranging from $h\nu = 20$~eV to 36~eV; an example of the corresponding band dispersion for this configuration can be seen on Fig.~3(f) (left side). For $k_x=\pm0.21(\pi/a)$, we clearly observe two bands (red and blue lines on Fig.~3(f)); the corresponding energy distribution curves (EDC), shown in Fig.~\ref{fig:FigV0}(a) for $h\nu = 23$ eV, have been carefully fitted using a sum of lorentzians multiplied by the Fermi function to determined the band positions as a function of $h\nu$. We also used a simple gaussian fit on the deeper band (band 1) to test that the result does not depend on the fitting procedure. The corresponding $k_z$ dispersions, displayed in Fig.~\ref{fig:FigV0}(b), show a clear periodicity as a function of $h\nu$. We observe a similar behavior for the band visible at $E_B=-0.25$~eV on Fig.~\ref{fig:FigV0}(a) (results not displayed).

To extract the inner potential value from the experimental band dispersion, we used the following method:
at \Ef\, the momentum in the direction perpendicular to the measurement is given by
\begin{equation}\label{eq:kz}
k_z=\frac{1}{\hbar} \sqrt{2m (E_{kin} cos^2{\theta}+V_0)}
\end{equation}
where $E_{kin}=h\nu-\phi$, with $\phi$ the work function ($\phi\approx4.5$~eV in our measurements) and $\theta$ the photoelectron emission angle.
The momentum value used in our study $k_x=\pm0.21(\pi/a)$ corresponds to an angle $3^{\circ} \leq \theta \leq 4^{\circ}$ for 36~eV~$\geq h\nu \geq $~20~eV,  \emph{i.e. } the measurement is close enough to normal emission to approximate $cos\:\theta\approx1$. We obtain the following relation:
\begin{equation*}
h\nu=\frac{\hbar^2 k_z^2} {2m}+\phi-V_0
\end{equation*}
At $\Gamma_n$, $k_z=\displaystyle\frac{2\pi}{c}n$ with $c=12.2443$~\AA\ \cite{Ueda2004}. Starting with a guess  of 11~eV  for $V_0$ (as typically found in heavy fermions materials) one can easily calculate from the above equation that the photon energies corresponding to ${\Gamma_3,\Gamma_4,\Gamma_5,\Gamma_6...}$ are $h\nu=$  \{2.5~eV, 9.5~eV, 18.5~eV, 29.5~eV...\}: the $k_z$ dispersion as a function of the photon energy should present extremum for all of the above energies. In other words, the band dispersion along the [001] direction must satisfy a sinusoidal function in which the period varies as $T_{n+1}=T_n+2$ with $h\nu$, where $T_0=7$~eV. Such a function can be described by:
\begin{multline*}
V(x)=A+B*cos \Bigg[ 
\pi+\frac{2\pi\:x}{T_0+2\:\big(p(x)-1\big)}\\
-\frac{2\pi\: \Big(x_0+T_0\:\big(p(x)-1\big)+\big(p(x)-1\big)\big(p(x)-2\big)\Big)}{T_0+2\:\big(p(x)-1\big)}   \Bigg]
\end{multline*}
where $x_0$ is the energy of the first extremum and
\begin{equation*}
p(x)=\bigg\lfloor \frac{1}{2}\Big(\sqrt{(T_0-1)^2+4(x-x_0)}-(T_0-3)\Big) \bigg\rfloor
\end{equation*}
If we now assume that $V_0=11$~eV~$+\delta V_0$, then $x_0=2.5$~eV~$+\delta V_0$. By using $V(x)$ as a fitting function (free parameters A, B, and $\delta V_0$) and taking the average for all the bands studied, we obtain a good estimate for the inner potential in \Ce: $V_0 =12 \pm 1$~eV. 

\begin{figure}
\makeatletter 
\renewcommand{\thefigure}{S\@arabic\c@figure}
\makeatother
\centerline{\includegraphics[width=.5\textwidth]{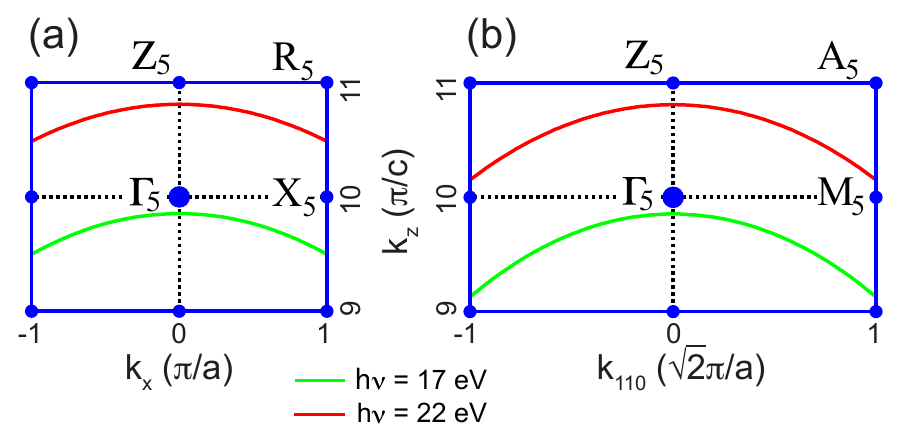}}
\caption{ARPES measurement at constant photon energy: 
In three-dimensional samples, ARPES data at different $k_{//}$ (\emph{i.e.} the momentum parallel to the surface) correspond to different values of $k_z$.
(a) and (b) show the \Ce\ Brillouin Zone in the (010) and (110) planes respectively. Green and red lines represent the probed \textbf{k} value for $h\nu=$17~eV and $h\nu=$22~eV respectively. }
\label{fig:Fighv}
\end{figure}

\subsection{ARPES measurement at constant photon energy.}

Using Eq.~\ref{eq:kz} with $V_0 =12$~eV, one can now calculate the correspondence between $(k_x,k_y)$ and $k_z$ for a given photon energy. The corresponding curves are displayed in Figure~\ref{fig:Fighv} for 17~eV (green curve) and 22~eV (red curve) in the (010) plane (a) and the (110) plane (b). 
Near to $(k_x,k_y)=(0,0)$, the experimental data shown in Fig.1 correspond approximatively to $k_z=1/7 (\pi/c)$ for $h\nu=17$~eV (a) and $k_z=4/5(\pi/c)$ for $h\nu=22$~eV (b) along the  $\Gamma \to Z$ symmetry line. 
As $(k_x,k_y)$ increases, one can see that while the probed $k_z$ variation represents  about one third of the $\Gamma_5\to Z_5$ distance in the (010) plane, it reaches up to 3/4 of the distance along the [110] direction. As a consequence, at the BZ boundary along the [110] direction, for  $h\nu=17$~eV (22~eV) the probed \textbf{k} almost reaches the symmetry point A (M), respectively.


\bibliography{HeavyFermions}




\end{document}